\documentclass[aps,prb,twocolumn,amsmath,amssymb,floatfix]{revtex4}
\usepackage{natbib}
\usepackage{tabularx}
\usepackage{bm}
\usepackage{euscript}
\usepackage{graphicx}
\usepackage{color}
\usepackage{amsfonts}
\usepackage{exscale}
\usepackage{amsbsy}
\usepackage{subfigure}
\usepackage{textcomp}
\usepackage{dcolumn}
\usepackage{amsmath}

\def\be{\begin{equation}}
\def\ee{\end{equation}}
\def\ba{\begin{eqnarray}}
\def\ea{\end{eqnarray}}

\def\C60{A$_x$C$_{60}$}

\begin{document}

\title{Kerr effect as evidence of gyrotropic order in the cuprates}
\author{ Pavan Hosur$^1$}
  \author{ A. Kapitulnik$^1$}
  \author{ S. A. Kivelson$^1$}
  \author{J. Orenstein$^{2,3}$}
  \author{ S. Raghu$^{1,4}$}
 \affiliation{$^1$Department of Physics, Stanford University, Stanford, CA 94305}
 \affiliation{$^2$Department of Physics, University of California, Berkeley, CA 94720}
 \affiliation{$^3$Materials Science Division, Lawrence Berkeley National Laboratory, Berkeley, CA 94720} 
\affiliation{$^4$SLAC National Accelerator Laboratory, Menlo Park, CA 94025 }
\begin{abstract}
The Kerr effect  can arise in a time-reversal invariant dissipative medium that is ``gyrotropic'', {\it i.e.} one that    breaks inversion ($\mathcal I$) and all mirror symmetries.  Examples of such systems include electron analogs of cholesteric liquid crystals, and their descendants, such as systems with chiral charge ordering.  We present arguments that the striking Kerr onset seen in the pseudogap phase of a large number of cuprate high temperature superconductors is evidence of chiral charge ordering.  We discuss additional experimental  consequences of a phase transition to a gyrotropic 
state, including the appearance of a zero field Nernst effect.  
\end{abstract}
\date{\today}

\maketitle

Key aspects of the physics of the cuprate high temperature superconductors (HTS) 
 remain controversial.  
A prominent feature of the normal state of the hole-doped cuprates is the depletion of spectral density at the Fermi energy which onsets below a somewhat imprecisely defined temperature, $T^\star$ - the enigmatic pseudogap crossover. 
There has been increasingly  strong 
 evidence that various forms of non-superconducting electronic order and incipient order occur as general features of at least the hole-doped cuprates in the pseudogap regime.\cite{Norman2004}  
 
In this paper, we will focus on one particular experimental probe of symmetry breaking - the  (polar) Kerr effect\footnote{The polar Kerr effect, which is what we will always mean by the Kerr effect, is measured at normal incidence.} - which has been measured with great sensitivity at optical frequencies.  In at least four families of cuprates -  underdoped YBCO-123\cite{KerrYBCO} and Hg-1201\cite{KerrHg1201}, optimally doped Bi-2201\cite{He2011}, and 1/8 doped LBCO\cite{KerrLBCO} -  there is a well defined  Kerr onset temperature, $T_K$, somewhere in the pseudogap regime.  
Above  $T_K$, the Kerr rotation angle $\theta_K$ 
is zero to within experimental accuracy, while it is non-zero for $T<T_K$.
This is 
indicative of a  symmetry breaking phase transition at $T_K$, despite the lack of clear and direct thermodynamic evidence of such a transition.

Since the Kerr effect is usually observed in association with magnetism, the most 
natural interpretation is  that
$T_K$ is associated with time-reversal symmetry ($\mathcal T$) breaking, in the same fashion as  an anomalous Hall effect would be. 
However, there are peculiar aspects of the Kerr measurements that cast doubt on this standard interpretation.  Firstly, the Kerr angle in all the cuprates measured to date cannot be ``trained" by 
cooling through the transition in an externally applied magnetic field.  
Secondly, recent measurements have shown that the sign of $\theta_K$ is the same for reflection on opposite surfaces\cite{KerrHg1201}.

 An alternative approach invokes the fact that 
  reflection from a 
 ``gyrotropic''\cite{Landau} material - {\it i.e.} one  with broken inversion ($\mathcal I$) and mirror symmetries -  also results in a non-zero 
 $\theta_K$, 
 even if  $\mathcal T$ is preserved\cite{Bungay1992,Mineev2010}.  Examples of such ``gyrotropic'' systems\cite{Landau} include electron analogs of cholesteric liquid crystals, and related systems with chiral density-wave ordering.

Consistent with the notion that a phase transition occurs at $T_K$, other  changes in physical properties have been documented at roughly the same temperature.  In the case of 1/8 doped LBCO,  $T_K$ 
coincides with $T_{{\rm co}}$\cite{KerrLBCO}, the temperature below which charge (stripe) order develops\cite{Tranquada2004,Tranquada2008,Tranquada2011}.  $T_{\rm co}$ also marks the onset of an anomalous zero-field Nernst effect\cite{NernstLBCO}, which we will discuss below.  In optimally doped Bi-2201, $T_K$ coincides with the temperature, $T_{pg}$, at which the pseudogap first opens, as detected directly in angle-resolved photoemission (ARPES)\cite{He2011} and NMR\cite{Kawasaki2010} experiments, and at which the relaxational dynamics abruptly change.\cite{He2011}  Finally,  in underdoped YBCO-123 for a range of doping, $T_K$ appears to track a characteristic temperature $T_H$\cite{Taillefer2011,Taillefer2012}, where an inflection point occurs in the Hall resistance.  Moreover, recently, it has been found that short-range (CDW) charge density wave order (detected in X-ray diffraction studies) first shows above the background at about the same temperature\cite{HaydenYBCO,RexsYBCO,XrayYBCO}.

In this paper, we suggest that $T_K$ represents a critical temperature in the pseudogap regime below which $\mathcal I$ and mirror symmetries are broken, resulting in gyrotropic order.  The observations of various forms of charge order in the  pseudogap regime, in conjunction with a non-zero $\theta_K$, implies that the charge order breaks $\mathcal I$, and can be thought of as signifying an electron cholesteric phase.     
It is important to stress that long-range chiral order can persist even if the density wave from which it is derived is only short-range correlated.
This proposal 
can naturally
rationalize the relation between $T_K$ and $T_H$, 
$T_{co}$, and $T_{pg}$, and the existence of a zero-field Nernst effect.
Specifically, our proposal offers a plausible candidate broken symmetry phase with a non-zero Kerr effect that  cannot be trained by a magnetic field, and that  has the same sign on opposite surfaces, although we do not at present have any compelling explanation of the peculiar ``memory effects'' (discussed below) seen in the Kerr measurements.

We familiarize the reader with the patterns of broken spatial symmetries which can give rise to the  Kerr effect in the absence of $\mathcal T$ breaking in Section I and with notions of gyrotropic order, familiar in the context of cholesteric liquid crystals, in Section II.
In Section III, we compute the gyrotropic response of various simple microscopic models of layered systems with stripe (smectic) or electron nematic order within a plane, and with a chiral ordering between layers.  In the remainder of the paper, we discuss the analysis of the Kerr experiments (Section IV), the relation with other probes of charge order in various cuprates (Section V), and finally, we speculate on the implications of these ideas for the understanding of the physics of the cuprates (Section VI).

\section{ Kerr
and Nernst effects
 in a gyrotropic medium  }

The Kerr effect measures the rotation of the plane of polarization of linearly polarized light when it is reflected off a surface
 (with $z$ defined to be the normal direction).  
 In the special case of a tetragonal system, a Kerr effect implies a non-zero relative phase acquired upon reflection by right- and left-circularly polarized light.
 In general, this requires an antisymmetric ``Hall-like'' component of the dielectric tensor.
 This can arise from an anomalous Hall effect in a system that breaks $\mathcal T$. A time-reversal symmetric system  can achieve the same if it is gyrotropic, {\it i.e.}, if it breaks all mirror symmetries. In either case, it is also necessary that the system be dissipative to produce a Kerr response.
 
 In a time-reversal invariant homogeneous medium, the dielectric tensor satisfies the reciprocity relations, $\epsilon_{ab}(\omega,\bm k)=\epsilon_{ba}(\omega,-\bm k)$;  thus, in the long wave-length limit, $\bm k\to \bm 0$, it is a symmetric tensor, and hence cannot give rise to a Kerr response.  However,   any odd in $\bm k$ contribution to $\epsilon$ must be antisymmetric:
\be
\epsilon_{ab}(\omega,\bm k) = \epsilon_{ab}(\omega) +i \gamma_{abc}(\omega)k_c+\ldots
\label{dielectric}
\ee
where $\epsilon_{ab}(\omega)=\epsilon_{ba}(\omega)\equiv\epsilon_{ab}(\omega,\bm 0)$, the gryotropic tensor\cite{Landau} $\gamma_{abc}(\omega)=-\gamma_{bac}(\omega)$ is the first derivative of $\epsilon_{ab}(\omega,\bm k)$ with respect to $k_c$, evaluated at $\bm k=0$, $k_c$ is the wavevector in the medium and $\ldots$ refers to higher order terms in powers of $\bm k$. It is easy to see that $\gamma_{abc}=0$  in a system with $\mathcal I$ or, for $a\neq c$,  if there is a reflection plane perpendicular to $\hat{ \bm e}_a$.   Thus, all these symmetries must be broken in order to obtain a gyrotropic Kerr response.  
If all these symmetries are broken, the medium is handed, and is said to be gyrotropic.

To be explicit, we consider the simple case of a tetragonal crystal in which the light is propagating along an axis, $z$, with $C_4$ symmetry.   Thus, for $a,b = x, y$, $\epsilon_{ab}(\omega)=\delta_{ab}\epsilon(\omega)$, the  gyrotropic tensor is defined by a pseudo-scalar amplitude, $\gamma(\omega) \equiv\gamma_{xyz}(\omega)=-\gamma_{yxz}(\omega)$, and $\gamma_{azz}=\gamma_{zaz}=0$.  In this case, the refractive indices for the two circular polarizations satisfy $n_{R(L)}(\omega) = \sqrt{\epsilon(\omega)\pm n_{R(L)}\omega\gamma(\omega)/c}$, which are distinct for $\gamma\neq0$. Note that  $\omega/c$ is the wavevector in vacuum which differs from the wavevector in (\ref{dielectric}) by a factor of the refractive index. The Kerr angle is then given by\cite{Bungay1992}
\begin{eqnarray}
\tan\theta_K & = & \mathcal Im\left[\frac{n_R-n_L}{n_Rn_L-1}\right] \nonumber\\
& = & - \frac \omega c {\mathcal Im} \left[\frac {\gamma(\omega)}{\epsilon(\omega) -1}\right].
\label{thetak}
\end{eqnarray}
to lowest order in $\gamma$. Clearly, a non-vanishing $\theta_K$ requires at least one out of $\epsilon(\omega)$ and $\gamma(\omega)$ to be complex and hence, $\epsilon(\omega, \bm k)$ to be non-Hermitian, which implies absorption.
The factor $\omega/c$ in this expression comes from the factor of $k$ in Eq. \ref{dielectric};  this implies an effect  is always small in proportion to the fine structure constant,  $\alpha=e^2/\hbar c$. 

A gyrotropic material also exhibits a  Faraday rotation, {\it i.e.} a non-zero $\theta_F$, the rotation of the plane of polarization of transmitted light:
\begin{eqnarray}
\theta_F &=& \frac{\omega d}{2c}\mathcal Re\left[n_R-n_L \right] \nonumber \\
             &=& \frac{\omega^2 d}{2 c^2} \mathcal Re[\gamma(\omega)],
             \end{eqnarray}
 where $d$ is the thickness of the material.               
The Faraday rotation can in principle be observed in thin films of a gyrotropic medium.

The transverse Peltier coefficient $\alpha_{xy}$ is a relative of the Hall conductance, $\sigma_{xy}$, and can be viewed similarly, as a probe of the same sort of spontaneously broken symmetries.  For simplicity, we focus on the DC response, {\it i.e.}  the limit $\bm k\to \bm 0$ and $\omega\to 0$ (in that order), where both $\sigma_{ab}$ and $\alpha_{ab}$ are real.  Time reversal symmetry requires $\sigma_{xy}=\sigma_{yx}$, {\it i.e.} a non-zero value of $\sigma_{xy}$ is equivalent to a misalignment of the principle axes of the conductivity tensor.  The presence of an $xz$ or $ yz$  mirror plane or a 4-fold rotational or screw axis is sufficient to imply $\sigma_{xy}=0$.  However, while either of these mirror symmetries would also imply $\alpha_{xy}=0$, a 4-fold rotational or screw symmetry, by itself, only implies that $\alpha_{xy}$ is antisymmetric: $\alpha_{xy} = -\alpha_{yx}$.  The difference between $\alpha_{ab}$ and $\sigma_{ab}$ is that the former involves the  correlation function between the electric and heat currents, while the latter involves only electric currents and so satisfies reciprocity relations.  Importantly, therefore, in a gyrotropic medium, where all reflection symmetries are broken, a non-zero $\alpha_{xy}$ is not forbidden.  
 This in turn can produce a zero field Nernst effect via $e_N = 
\alpha_{xy}/\sigma_{xx}$.

\section{ Electron cholesteric order  }  
Liquids and (cholesteric) liquid crystals with the requisite $\mathcal I$ breaking needed for a gyrotropic response are generally formed with some molecular constituents which are intrinsically chiral.  Thus, $\mathcal I$ is explicitly, rather than spontaneously broken.  For instance, when a small concentration of chiral molecules is added to a nematic liquid crystal, a new term in the Landau-Ginzburg (LG) free energy density for the nematic director field, $\vec n$, is induced of the form
\be
\delta {\cal F} = -\chi (\vec n\cdot\vec \nabla\times\vec n)
\label{nematicfreeenergy}
\ee
where the sign of $\chi$ is determined by the handedness of the chiral molecules, and its magnitude is at least roughly proportional to their concentration.\cite{Chaikin}  In turn, the presence of this first-derivative term  implies that for any non-zero $\chi$,  a uniform nematic phase will be replaced by a cholesteric phase in which the nematic director forms a helical texture of the form 
\be
\vec n(\vec r)=n_0[\cos(\pi Qz),\sin(\pi Qz),0]
\label{nematichelix}
\ee
 with a pitch, $1/Q$, proportional to $\chi$.  (Recall, $\vec n$ is a headless vector, $\vec n\equiv -\vec n$, so $1/Q$ is the period of the helix.) 
 Even in the disordered phase, the presence of such a term implies that $<\vec n\cdot\vec \nabla\times\vec n>\propto \chi\neq 0$, so the chirality of the constituents manifests as the non-vanishing helicity of the nematic fluctuations.  $\chi$ is a psuedo-scalar quantity which characterizes the chirality of the fluid;  one would generically expect a fluid with a small non-zero $\chi$ to exhibit  a non-vanishing value of  $\gamma \propto \chi$.

Turning to the electron fluid in solids, if the crystal structure itself is non-centro-symmetric and is appropriately lacking in mirror symmetries, this will generically give rise to an extrinsic non-zero value of both $\gamma$ and $\chi$.  On the other hand, in a centro-symmetric crystal, a non-zero value of $\chi$ can develop as a consequence of spontaneous symmetry breaking.  
It is important to note that, even in the case in which  no other order parameter is needed, it may be useful to think of the pseudo-scalar order parameter, $\chi$, as representing the residual symmetry breaking remaining upon melting of a chiral nematic.  
\section{ Model problems  }
To get a feeling for the way in which various patterns of symmetry breaking lead to gyrotropic behavior, we have analyzed a number of simple model problems, ranging from the phenomenological to the microscopic. All of our microscopic models are of perfect conductors (no quasi-particle scattering) so in each case $\gamma$ is real;  to produce a Kerr effect, some source of dissipation must be present.
\begin{figure}[t]
\begin{minipage}{0.99\linewidth}
\includegraphics[width=\linewidth]{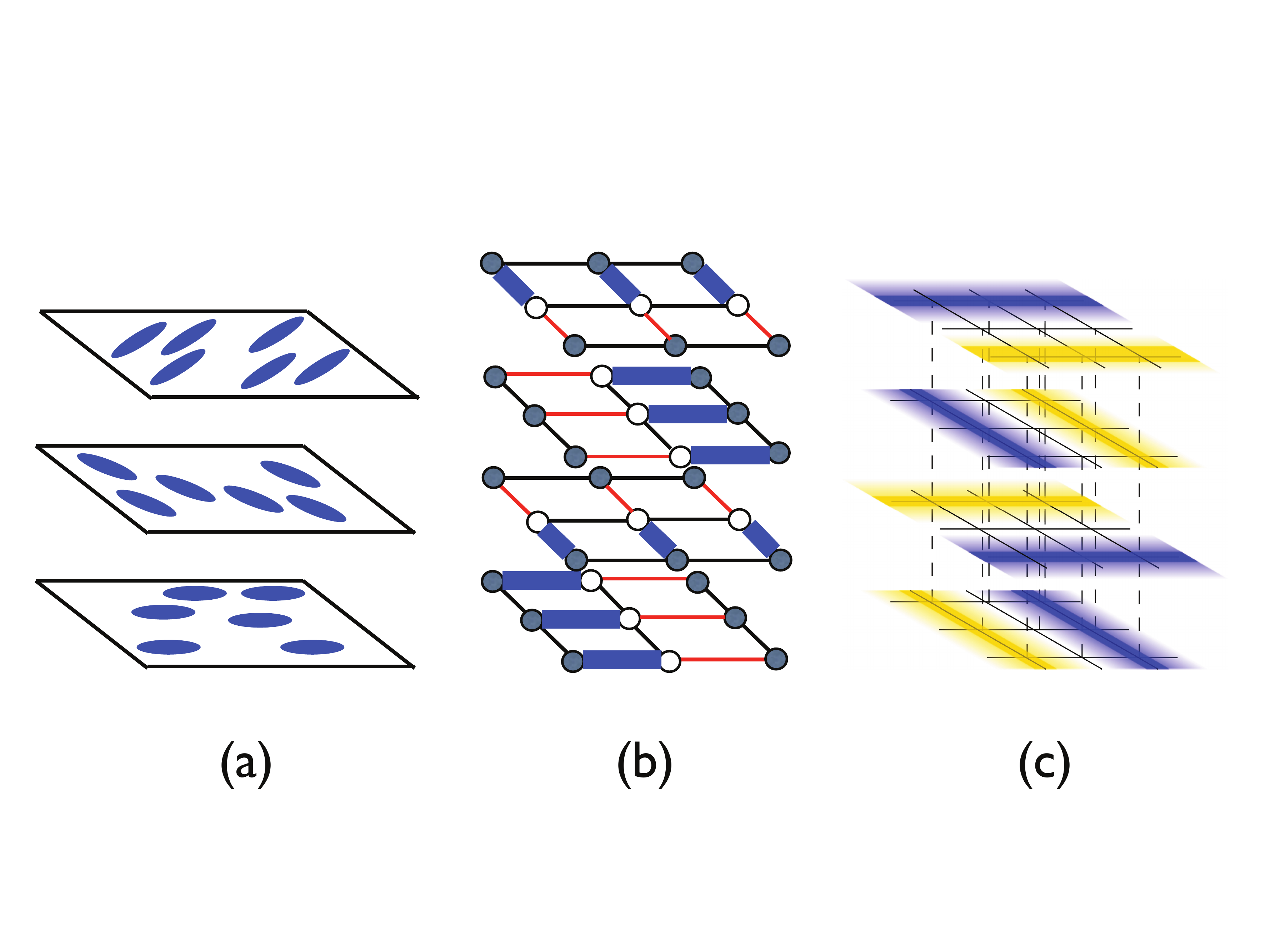}
\end{minipage}
\caption{ Schematic representation of:  a)  A period 3 cholesteric phase, where the  ellipses  represent a distorted Fermi surface in each plane, and  the orientation of the major axis represents the nematic director.  b)  A chiral pattern of stripe order in a tetragonal background.  The filled (empty) circles represent sites with excess (reduced) charge density, while the heavy blue (thin red) lines represent strong (weak) bonds.  This state has period 2 unidirectional (stripe) order in each plane, with a 4-plane unit cell.  Importantly, the stripes break a mirror symmetry as they are neither site centered (which would correspond to vanishing bond-charge modulation) nor bond centered (which would correspond to vanishing site-charge modulation).  c)   Another 4-plane chiral stripe ordered state, but now with period 3 mirror-plane breaking stripe order in each plane. The broad blue, narrow black, and broad yellow lines represent, respectively, lines of excess, average, and reduced charge density.}
\label{gyrofig}
\end{figure} 
\subsection{ A phenomenological model}   To begin with, we consider a phenomenological model of a chiral nematic in a layered material, where the nematic order parameter has the form shown in Eq.\ref{nematichelix} with integer $z$ labeling the planes, and with a commensurate pitch $1/Q>2$.    In each plane, we imagine a conductivity tensor of the form $\sigma_{ab}(z)=\sigma_0 + \sigma_1  n_a(z) n_b(z)$.  We further imagine a coupling between currents in neighboring planes (related to the drag), so that
\be
j_a(z) = \sigma_{ab}(z)E_b(z) + D_{ab}(z)[j_b(z+1)+j_b(z-1)]
\ee
where we have adopted summation convention, $j_a(z)$ and $E_a(z)$ are, respectively, the current and electric field in the $a$ direction in plane $z$, and the drag tensor is given by 
$D_{ab}(z)=D_0 + D_1  n_a(z) n_b(z)$.  Assuming the relation $\epsilon=1+4\pi\sigma/i\omega$ and the anisotropic part of the drag $D_1\ll 1$, we can compute the dielectric and gyrotropic responses: 
\begin{eqnarray}
\epsilon_{xx}  = \epsilon_{yy}  &=& 1+\frac{4 \pi}{\omega}\frac{(2\sigma_{0}+n_{0}^{2}\sigma_{1})\cos(2\pi Q)}{1-2D_{0}\cos(2\pi Q)}\nonumber\\
\gamma  &=&  \frac{4 \pi}{i\omega}\frac{n_{0}^{4}D_{1}\sigma_{1}\sin(2\pi Q)}{(1-2D_{0})\left[1-2D_{0}\cos(2\pi Q)\right]^{2}}
\label{dragresult}
\end{eqnarray}
Notice that, as it should, $\gamma$ changes sign upon $Q\to-Q$ ({\it i.e.} upon flipping the chirality).

\subsection{Period 3 electron cholesteric}
The simplest microscopic model of a gyrotropic system describes electrons in non-centro-symmetric crystals with $\bm k\cdot\bm\sigma$ type spin-orbit coupling.  The Kerr response in such a system was computed and shown to be non-zero recently in Ref. \onlinecite{Mineev2010}.   
However, with the cuprates in mind, we wish to consider model problems of quasi-2D (layered) materials in which the non-zero Kerr response is a consequence of spontaneous symmetry breaking (charge ordering).

First, we study the simple  electronic analog of a cholesteric liquid crystal shown in Fig. \ref{gyrofig}(a).  Each layer consists of spinless electrons (the spin degree of freedom is a mere spectator in the analyses to follow) forming an electron nematic, and neighboring layers are coupled via a hopping integral $t_{\perp}$:
\begin{eqnarray}
&&H =  \sum_{\bm k} E(\bm k;z)  \psi^{\dagger}_{\bm k, z}   \psi_{\bm k, z} - t_{\perp} \sum_{\bm k, z} \left( \psi^{\dagger}_{\bm k,z} \psi_{\bm k, z + 1} + h.c. \right) \nonumber \\
&&E({\bm k};{z})  = \frac{1}{2m } \left[ k^2+ \left[{\bm k}\cdot{\bm n}(z) \right]^2 \right] - E_F
\label{3layernematic}
\end{eqnarray}  
where $\bm k$ is the in-plane momentum, $z$ is 
an integer which indexes the layer,
and ${\bm n}$ is given by  Eq. \ref{nematichelix} with $Q=\tilde\chi/3$ where $\tilde \chi=\pm1$ determines the chirality.
 In the limit where 
 $|t_{\perp}| \ll E_F \equiv k_F^2/2m$, a closed form expression for the gyrotropic tensor can be obtained using the Kubo formula,  treating the interlayer tunneling as a perturbation:  
\begin{eqnarray}
\gamma(\omega) 
&\simeq&\frac {e^2}{\hbar \omega} \chi \left(\frac {t_\perp}{E_F}\right)^2 F\left(\frac{\hbar \omega}{E_F},|\chi|\right) 
\label{3layerresult}
\end{eqnarray}
where $\chi=\tilde \chi n_0^4$ and 
\be
F(x,y)= \left(\frac 1 x \right)^2\sqrt{\frac 3 {x^2 -3|y|/4}}.
\ee
This expression is valid in the limit 
$n_0^2 \ll 1$, $|t_{\perp}| \ll E_F n_0^2$, and $|t_{\perp}|\ll \hbar\omega$. 
The gyrotropic response of the system vanishes in the absence of interlayer tunneling, as the chirality is ill-defined for decoupled planes.  
\subsection{Chiral stacking of commensurate stripes}
We now turn to models in which a chiral CDW 
produces a gyrotropic effect. 
Specifically, in a layered tetragonal crystal, we consider  an ordered phase consisting of a unidirectional CDW (stripes) in each plane with a direction of propagation (which one can think of as the nematic component of the CDW) that rotates by $\pi/2$ between neighboring planes.  
Moreover, 
motivated by  experiments in the cuprates, we consider 
a state with a period 4 in the $z$ direction.  In order to give rise to a gyrotropic effect, the overall stripe order must break  mirror symmetry through the planes perpendicular to the direction of propagation.

As a first example, we consider period 2 stripe order shown in Fig. \ref{gyrofig}b.
The corresponding mean-field Hamiltonian is
\ba
&&H=H_0 +H_{CDW} \label{HCDW}\\
&&H_{CDW}= \sum_{\bm r} \left \{\Phi(\bm r) \rho({\bm r}) +\Delta_x(\bm r)T_x(\bm r)+\Delta_y(\bm r) T_y(\bm r)\right\}, 
\nonumber
\ea
where ${\bm r}=(x,y,z)$ and  $x$, $y$, $z$ are the integer valued coordinates of the lattice sites, $H_0$ is the band-structure of the undistorted tetragonal lattice, 
 the ``site charge density''  and ``bond-charge density'' are, respectively,
\be
 \rho({\bm r}) =  c_{\bm r}^\dagger c_{\bm r} \ \ {\rm and} \ \  T_a({\bm r}) =  \left[c_{\bm r}^\dagger c_{\bm r+\hat e_a} + H.C.\right]\label{nanddelta},
\ee
and 
$c^\dagger_{\bm r}$ creates an electron on site ${\bm r}$.  
We  parameterize the order parameter as follows:
\ba
\Phi({\bm r}) =\Phi &&\left[\cos\left(\frac{\pi (z-z_0)}{2}\right)\cos\left(\frac{2\pi x}N+\theta_x\right) \right . \nonumber \\
&&\left . 
+\tilde \chi\sin\left(\frac{\pi (z-z_0)}{2}\right)\cos\left(\frac{2\pi y}N+\theta_y\right)\right],
\nonumber \\
&&\Delta_x (\bm r)= \Delta \frac{ (1+e^{i\pi z})}2\cos\left(\frac{2\pi x} N\right), \nonumber \\ &&\Delta_y(\bm r)= \Delta \frac {(1-e^{i\pi z})} 2\cos\left(\frac{2\pi y} N\right), 
\ea
where $N=2$ is the commensurability,  and $\tilde\chi=\pm1$ is the chirality of the gyrotropic phase. For later use, we have introduced CDW phases $\theta_a$ and $z_0$, which are redundant (and can be set equal to zero) for the case of  $N=2$. In this case, for  purely site-centered stripes ($\Delta=0$) or purely bond-centered stripes ($\Phi=0$), the state is mirror symmetric, and so non-gyrotropic.  For stripes that are neither site- nor bond-centered, we can associate a preferred direction with each plane by drawing an arrow from any site with maximal charge density along the direction of the strongest bond radiating from it, {\it i.e.} from the filled sites along the thick-blue bonds in Fig. \ref{gyrofig}b.  
  For $\chi=1$, it is clear from the figure that this direction rotates by $\pi/2$ in a counter-clockwise direction in passing from any plane to the plane above, $z\to z+1$, defining a right-handed helix with period 4.

 In the limit 
 of weak inter-plane coupling 
and weak density wave order, $t_\perp$, $\Phi$, and $\Delta \ll \mu$, an estimate of the gyrotropic tensor can be obtained from a symmetry analysis, resulting in an expression of the same form as in Eq. \ref{3layerresult} with
\be
\chi=\tilde \chi \left[\Phi^2\Delta^2/E_F^4\right]
\ee
and the dimensionless function $F$ can, in principle, be computed from the Kubo formula given  an explicit form of $H_0$.

Finally, we consider a model in which there are higher-order commensurate stripes in each plane, 
corresponding to the pattern of CDW ordering shown in Fig \ref{gyrofig}(c) for the case of commensurability 3.
Now, even for a pure site-charge density wave, as long as the
antinodes of the CDW do not fall on the lattice sites, reflection
symmetry about a plane perpendicular to the CDW ordering vector in each plane is broken, with no need for an explicit bond density wave.  
Thus, we consider a system described by $H$ in Eq. \ref{HCDW} with $\Delta_a(\bm r)=0$,
where $N \geq 3$ is the commensurability 
and an integer-valued $z_0$ ensures that the stripes are either along $x$ or along $y$ in every plane.
The ordering vectors in this model are $( 1/N,0, 1/4)$ and $(0, 1/N,1/4)$.
(For $N=3$, these 
are  close commensurate approximants to the recently detected charge ordering peaks seen in YBCO, which will be discussed below.)  

For $\theta_y= n\pi /N$ 
  $xz$ mirror symmetry is preserved, while for $\theta_x= n\pi /N$ $yz$ mirror symmetry is preserved;  in either case, there will be no gyrotropic response. For all other $\theta_x$ and $\theta_y$, mirror symmetries are broken, and a gyrotropic response is expected.   $\gamma$ must be invariant under any transformation which preserves the chirality:   1) translation in $a=x$ or $y$ takes  $\theta_a\to \theta_a + 2\pi/N$, 2) translation in $z$ takes $z_0\to z_0+1$ 3) $C_2$ rotation about the $x$ axis takes  $z_0\to-z_0$ and $\theta_y\to\pi-\theta_y$ 4) $C_2$ rotation about the $x=y$, $z=0$ line interchanges $\theta_x$ and $\theta_y$ and takes $z_0\to 1-z_0$. Reflection must invert the chirality, taking $\gamma\to -\gamma$. $xy$ reflection takes $z_0\to-z_0$. All other point group transformations can be obtained as products of two or more of the above operations. Combining these considerations with an expansion in powers of $\Phi$ and $t_\perp$, we 
    again obtain an expression of the same form as in Eq. \ref{3layerresult} with
\be
\chi =\tilde \chi 
\begin{cases} \phi^{2N}\sin(N\theta_x)\sin(N\theta_y) & \text{for odd }N\\
\phi^{4N}\sin^2(N\theta_x)\sin^2(N\theta_y) & \text{for even }N\end{cases}
\ee
where $\phi\equiv \Phi/E_F$ and, again, a dimensionless function $F$ which could be computed for a given $H_0$.

\section{Analyzing Kerr experiments } 
There are various possible broken symmetries that can lead to a Kerr signal.  
 In general, various protocols, including examining the consequences of different thermal histories, 
 and introducing symmetry breaking external fields, can be used to distinguish among these possibilities to a considerable extent.  Spatial transformations, such as ``flipping" the sample ({\it i.e.} rotating it by $\pi$ about the $x$ or $y$ axis) can also be used to discriminate among various broken symmetries.  Broadly speaking, it is possible to classify systems that produce a Kerr signal as 1) those that break $\mathcal T$ and 2) those that preserve it.  In case 1, there is a subclassification based on 1a) the presence or 1b) absence of an anomalous Hall response.  We discuss each of these cases separately.    

In case 1a) the  order parameter is an axial vector, such as the magnetization $M_z$, which couples linearly to an external magnetic field, $H_z$, applied in the $z$ direction.  When such a system is cooled in zero field, the sign of the Kerr signal is expected to vary randomly in sign (and in the case in which there are multiple chiral domains, also randomly in magnitude) between different thermal cycles.  
However, if the sample is slowly cooled through $T_K$ in a positive applied field, $H_z > 0$,  this should align the magnetic domains.   Consequently, the Kerr signal measured upon subsequent heating in zero field will exhibit a fixed  positive Kerr angle, which vanishes only as $T\to T_K$.  Moreover, if the identical procedure is repeated with $H_z  < 0$, the reversed Kerr angle 
must be observed.  
For instance, the same apparatus used to measure the cuprates was first used to study the transition metal oxide ferromagnet, SrRuO$_3$\cite{Xia2006a}, where 1) zero-field-cooled samples exhibited  random magnitude and sign of the Kerr angle below the Curie temperature, and 2) training of the Kerr angle occurred in field-cooled samples, with $\theta_K \sim  2\times 10^{-2}$ at low temperatures.  
A second example  is 
the spin-triplet superconductor Sr$_2$RuO$_4$\cite{Xia2006},
in which $\theta_K\sim 10^{-8}$. Here $T_K=T_c$, the superconducting transition temperature, so the observed field-training of $\theta_K$
unambiguously 
identifies the superconducting transition  
as being $\mathcal T$ breaking.

In case 1b) the order parameter is a diagonal, rank-2 pseudotensor\footnote{Off-diagonal components of $\theta_{ab}$ are invariant under reflections in various vertical mirror plane and are therefore do not produce a Kerr response} $\theta_{ab}$, that couples to the product $E_a H_b$.  Such a system also breaks $\mathcal I$ but preserves the product  $\mathcal T \mathcal I$.  If the system only has a non-zero $z$-component $\theta_{zz}$, then the Kerr response can be trained by $H_z$.  On the other hand, if only in-plane components $\theta_{xx}, \theta_{yy}$ are present, the Kerr angle in such a system cannot be trained by $H_z$.    Furthermore, the Kerr effect is even under flipping the sample in such systems\cite{Orenstein2011}.

In case 2, $\mathcal T$ is preserved, and extrinsic strains, or effects of the sample geometry can act as symmetry breaking fields on the chiral domains, which can often lead to a fixed Kerr signal for subsequent cool-downs.  More importantly, while it is possible  that the application of a field $H$ at some points in the thermal cycle may affect the outcome, the result should be independent of the {\it sign} of the applied magnetic field.

In all the cuprates measured to date, the Kerr signal appears to have a fixed sign upon multiple cool-downs.  Moreover, even if the sample is cooled through $T_K$ in the presence of a field, $-H$, the sign of the Kerr signal typically remains unchanged.
This ``memory effect'' implies that the Kerr onset in the cuprates is not directly associated with $\mathcal T$ breaking.  
It is possible, however, to imagine a memory effect of extrinsic origin;  for instance, if there were small ferromagnetic inclusions in the sample, they could induce a fixed sign of the Kerr effect. 
 Since a non-zero Kerr effect has, by now, been seen in a wide variety of different cuprate crystals and films, with different origins, such an extrinsic origin of the memory effect 
is unlikely.  Moreover, recently, the effect of flipping the crystal has been tested on crystals of YBCO, Bi2201, and LBCO, and found to have no effect on the sign of the Kerr angle,  consistent with a pseudo-scalar order parameter.  This experiment rules out any interpretation associated with a bulk anomalous Hall effect.

The magnitude of the dimensionless Kerr angle itself has some significance, and introduces ``reality checks" on proposed theories.  If the Kerr rotation arises in such systems from gyrotropic order, $\theta_K$ generically is small, since it is proportional to  
 the fine structure constant $\alpha$.  In a quasi-2D material, a second small parameter, 
 the anisotropy  $(t_\perp/E_F)^2$, further reduces the expected magnitude, as is clear from Eqs. \ref{thetak} and \ref{3layerresult}.  
 An estimate of the latter quantity can be obtained from the ratio of temperature derivatives of $\rho_{ab}$ and $\rho_c$, evaluated at or above room temperature\footnote{The resistivities themselves are highly temperature and frequency dependent and their ratios are determined by additional factors than the underlying quasi-two dimensionality.  However, since they are linear functions of temperature at high temperatures, the ratio of their slopes 
 above room temperature provide a roughly temperature-independent estimate of the anisotropy.}.  
 Together, these two factors can roughly account for the magnitude of $\theta_K$ observed in the cuprates.
 By contrast, any mechanism that produced an anomalous Hall  response in a single plane might be expected to give values of $\theta_K$ several orders of magnitude larger than those observed in the cuprates, unless that Hall response itself were parametrically small(see for example Ref. \onlinecite{Tewari2008}).

\section{Kerr results in context}
A bewildering number of crossover phenomena - including putative transitions - have been identified in various cuprates in different regimes of temperature and doping.  It is clearly important to correlate  the various phenomena, and to identify 
  the common features of the electronic structure that lead to them.

{\bf LBCO near 1/8 doping:}  LBCO is an anomalous cuprate both in the sense that it has a much suppressed superconducting $T_c \approx 4K$, and that it has a subtly distorted crystal structure (the so-called LTT phase) which stabilizes long-range correlated, static stripe order\cite{Tranquada2004,Tranquada2008}.  However, by the same token, this makes possible declarative statements about the relation of various other observed phenomena to charge and spin density wave order to a much greater extent than in other cuprates.

There is a transition at $T_{co}=52K$ to a charge ordered state with primary ordering vectors $Q_x=(q,0,1/2)$ and $Q_y=(0,q,1/2)$ with $q\approx 0.23$ close to, but not equal to the commensurate value $1/4$\cite{Hucker2011,Wilkins2011}.  The in-plane correlation length (inferred from the width of the peaks in either X-ray or neutron scattering) is around 100 lattice constants, while perpendicular to the planes, the correlation length is on the order of one unit cell.  As the LTT structure has two planes per unit cell, the $1/2$ in the c-direction implies a periodicity in that direction of four planes.  Within the unit cell, the order is known to correspond to x-directed charge stripes (ordering vector $Q_y$) in the first and third plane, shifted by half a period relative to each other, and y-directed charge stripes (ordering vector $Q_x$) in the second and fourth plane, again shifted by half a period.  Spin ordering detected in neutron scattering onsets at a distinctly lower temperature, $T_{so}=42K$, with related ordering vectors $(1/2)(1\pm q,1,L)$ and  $(1/2)(1,1\pm q,L)$.  NMR\cite{Grafe2010} and $\mu$SR \cite{Savici2005} studies confirm the onset of magnetic order with around the same onset temperature.

The Kerr signal is observed to onset at the charge ordering temperature, $T_K \approx T_{co} > T_{so}$\cite{KerrLBCO}. Previous discussions of the stripe order have primarily focused on its unidirectional (electron-nematic) character within each plane, rather than its (short-range correlated) interplane structure.  While the nematic order has the interplane period 2, inherited directly from the LTT structure itself, 
 the period 4 density wave order described above is the same as in our simple models, and so  can exhibit gyrotropic behavior. 
 Thus, we suggest that the Kerr signal arises from the pseudo-scalar chiral component, $\chi$, of  3D charge stripe order.

It is to be expected that the charge ordering triggers significant changes in the electronic structure, which in turn can produce significant changes in transport properties.  This has been confirmed in multiple experiments.  In particular, the Hall resistance, which is positive and an increasing function of decreasing temperature, begins to drop rapidly toward zero (and can ultimately  change sign at low $T$) at $T_{co}$\cite{Adachi2011}, and similar behavior is seen in the thermopower\cite{Li2007}.  

Recently, an apparent ``anomalous'', zero field Nernst effect\cite{NernstLBCO} has been observed.  
At face value, this effect seemed to imply that $\mathcal T$ is broken at $T_{co}$.  However, 
it could alternatively reflect the same gyrotropic order we have proposed as the origin of the Kerr effect.
This interpretation of the anomalous Nernst data can be tested by a variety of protocols. For instance, $e_N$ should not change sign on flipping the sample in a gyrotropic system.

{\bf Underdoped YBCO:}  The Kerr onset in YBCO has been tracked 
over the entire range of doping up to and including optimal doping.  $T_K$ is found to decrease strongly with increasing doping, more or less following the trend of the generally accepted pseudogap crossover, $T^\star$, although $T_K$ is systematically smaller than $T^\star$ by up to 100K.  However, a number of recent experiments have independently identified a characteristic temperature, which appears to correlate quantitatively with $T_K$.  

In transport, a temperature $T_H$ (defined as an inflection point in $R_H$ vs $T$) has been identified 
with an inflection point in the Hall coefficient, $R_H$, 
as a function of temperature\cite{Taillefer2011,Taillefer2012}.  It has been argued that this temperature marks the onset of the Fermi surface reconstruction that will lead, at lower temperature, to the observed sign change of the Hall resistance and, ultimately, to the existence of the electron pockets inferred from quantum oscillation experiments\cite{Proust2007,Proust2007a,Sebastian2010}.  
Similar crossover behavior has been observed in the thermopower\cite{Chang2010}, with the same characteristic temperature.
Both qualitatively and quantitatively, $T_H$ appears to correlate well with $T_K$.

A recent high field NMR study has identified a magnetic field induced transition to a statically ordered charge density wave phase at low temperatures\cite{NMRYBCO}.  The same experiments find no evidence of magnetic order at any temperature in the field-range explored.  Together, these results strongly support a correlation between  charge ordering phenomena and the most significant features of the Fermi surface reconstruction.

More recently still, two  X-ray scattering experiments have revealed the existence of significant peaks in the structure factor at wave-vectors $Q_x=(q,0,1/2)$ and $Q_y=(0,q,1/2)$ with $q\approx 0.31$ in underdoped YBCO with doped hole concentration $x$ near 1/8\cite{HaydenYBCO,RexsYBCO,XrayYBCO}.  As in the case of LBCO, there are two Cu-O layers per crystalline unit cell, so the interplane ordering vector 1/2 implies four planes per unit cell.  The width of the peaks in-plane implies correlation lengths of up to $\xi_{ab} \sim 20$ lattice constants, but the correlation length in the interplane direction is of order one unit cell.  That the associated CDW order is fluctuating, not static, is inferred from the observed strong $T$ dependence of $\xi_{ab}$, and from the fact that no evidence of associated static electric quadrupole ordering is seen in NMR studies of the same materials.\cite{NMRYBCO}   

The observation that the peaks at $Q_x$ and $Q_y$ have approximately equal intensity, width (correlation length) and incommensurability, $q$, is somewhat surprising, given the strongly orthorhombic character of the electronic structure of YBCO.  The similarity of the pattern of ordering vectors to those seen in LBCO (with a somewhat different value of $q$) is possibly suggestive that here, too, the charge order corresponds to alternating planes with $x$ and $y$ directed charge stripes.   Evidence of a nematic character of the electronic state has also been inferred from measurements of the anisotropy of the Nernst effect (antisymmetrized with respect to the magnetic field). Although YBCO is orthorhombic, and hence always has a non-zero Nernst anisotropy, an anomalous increase of the anisotropy occurs in the pseudogap regime, which has been associated with spontaneous symmetry breaking in the electronic system.  While the onset temperature of this effect is somewhat difficult to determine, it seems to have the same trends with doping as $T_K$.  
It will require further experiments to disentangle any possible symmetric in magnetic field portion of the Nernst response from an admixture of the thermopower due to misalignment of the leads.

{\bf Optimally doped Bi2201:}  A coordinated study of the Kerr effect, pump and probe optical reflectivity, and ARPES was carried out on an optimally doped crystal of Bi2201.\cite{He2011}  $T_K$ was found to coincide with the temperature at which an antinodal pseudogap first begins to develop in the ARPES spectrum, and at which sharp changes in the relaxational dynamics could be identified in pump and probe experiments.  This observation establishes a close relation between the ordering tendencies which give rise to the Kerr effect, and the fundamental physics of the pseudogap.  $T_K$ also corresponds well to the value of $T^\star$ inferred from the onset of $T$ dependence of the Knight shift observed in NMR studies\cite{Kawasaki2010} of similar crystals, giving bulk confirmation of the same relationship.  It should further be noted that the same NMR studies find no evidence of magnetic order to the lowest temperatures, even when superconductivity is quenched by the application of a high magnetic field.  Conversely, some preliminary evidence of CDW order - which possibly breaks $\mathcal T$ and/or $\mathcal I$ - has been obtained from inelastic X-ray scattering in this material.\cite{Bonnoit2012}

\section{ Discussion}
Concerning electronic order in the pseudogap regime of the underdoped cuprates, we propose the following:  1)  That there is a phase transition at $T_K$, at which $\mathcal I$ and mirror symmetries are broken. 2)  That this transition reflects a local tendency toward CDW formation within the planes, but does not require long-range CDW order.
If confirmed, the existence  of a non zero value of the pseudo-scalar $\chi\neq 0$ is interesting in its own right.  However, by itself, this order is unlikely to play a central role in the physics of the cuprates - at best it is a useful {\it indicator} of important changes in the electronic structure which occur in the pseudogap regime.  In the first place, it reflects subtle correlations between neighboring Cu-O planes, whereas the essential physics of the cuprates is clearly associated with the much larger intra-plane electronic 
interactions.  

While it seems likely that the gyrotropic order derives from an underlying tendency toward CDW formation, the origin of the CDW formation and its relation to the pseudogap remain open questions.  For instance, stripe order could reflect a still more basic tendency toward incommensurate spin-density-wave\cite{Zaanen1989,Zachar1998} or d-density wave\cite{Eun2012} formation.
Moreover, all independent measures of CDW order in the cuprates show that the strength of the ordering tendency is strongly peaked in the neighborhood of $x\sim 1/8$, while the pseudogap phenomena apparently grow monotonically stronger with decreasing $x$.  All this is suggestive that the stripe order, and its avatars, electron nematic and cholesteric order, are probably all {\it consequences} of more fundamental electronic correlations.

Finally, there are at least two  noteworthy 
experimental observations that are not immediately accessible  in the present framework.
  Firstly, the origin of the memory effect above T$_K$ remains a mystery.  Phenomenologically,  one needs to imagine that there is an effective weak symmetry breaking field in each crystal that determines the chirality of the ordered phase upon cooling below $T_K$, but the origin of this field is, presently, unclear.  
Secondly, evidence of a translation symmetry preserving magnetic  phase\cite{Varma1997} which onsets below the pseudo gap $T^\star$ (which, at least in the case of YBCO, is distinctly larger than $T_K$) has been obtained from magnetic neutron scattering studies in YBCO\cite{Bourges2006} and Hg-1201\cite{Greven2010}.  
Evidence has also been adduced\cite{Eun2012} of a transition to a d-density wave state at $T^\star$.  Moreover, the existence of distinct thermodynamic signatures of transitions at $T^\star$ and $T_K$ has been reported in recent resonant ultrasound measurements\cite{Shekhter2012,Bhattacharya1988,Bhattacharya1988a} in YBCO.  The relation between $T^\star$ and $T_K$ is presently an important open question.

\acknowledgements{  We acknowledge important conversations with Peter Armitage, Sudip Chakravarty, Eduardo Fradkin, Alexander Fried, Ruihua He, Daniel Podolsky, Chandra Varma, and especially John Tranquada.  This work was supported in part by the Office of Basic Energy Sciences, Materials Sciences and Engineering Division of the U.S. Department of Energy under Contract Nos. DE-AC02-05CH11231 (JO), and AC02-76SF00515 (AK, SK, SR), and the Alfred P. Sloan Foundation (SR). }

\bibliography{kerr_sri}

\end{document}